\documentclass[12pt]{iopart}
\usepackage{graphics,epsfig,color,verbatim,ulem}

 \newcommand{\vk}{{\mathbf{k}}}
 \newcommand{\vq}{{\mathbf{q}}}

\newcommand{\trace}{\mbox{Tr}}

\begin{document}

\title{Coherence-incoherence crossover in the normal state of iron-oxypnictides and importance of the Hund's rule coupling}

\author{K. Haule and G. Kotliar}

\address{
Department of Physics, Rutgers University,
Piscataway, NJ 08854, USA}
\ead{haule@physics.rutgers.edu}

\begin{abstract}{ 
    A new class of high temperature superconductors based on iron and
    arsenic was recently discovered \cite{Kamihara:2008}, with
    superconducting transition temperature as high as 55$\,$K
    \cite{Sm}.  Here we show, using microscopic theory, that the
    normal state of the iron pnictides at high temperatures is highly
    anomalous, displaying a very enhanced magnetic susceptibility and
    a linear temperature dependence of the resistivity.  Below a
    coherence scale $T^*$, the resistivity sharply drops and
    susceptibility crosses over to Pauli-like temperature
    dependence. Remarkably, the coherence-incoherence crossover
    temperature is a very strong function of the strength of the
    Hund's rule coupling $J_{Hund}$.  On the basis of the normal state
    properties, we estimate $J_{Hund}$ to be $\sim 0.35\,$eV.  In the
    atomic limit, this value of $J_{Hund}$ leads to the critical ratio
    of the exchange constants $J_1/J_2 \sim 2$.  While normal state
    incoherence is in common to all strongly correlated
    superconductors, the mechanism for emergence of the incoherent
    state in iron-oxypnictides, is unique due to its multiorbital
    electronic structure.
  }
\end{abstract}

\maketitle

The unusually high superconducting critical temperatures in
iron-oxypnictides together with unusual normal state properties, which
do not fit within the standard framework of the Fermi liquid theory of
solids, place the iron pnictides in the broad category of strongly
correlated superconductors, such as $\kappa$ organics, cerium and
plutonium based heavy fermions, and cuprate high temperature
superconductors. In all these materials, superconductivity emerges in
close proximity to an incoherent state with unconventional spin
dynamics, that cannot be describe in terms of weakly interacting
quasiparticles. Describing the normal state of these materials, is one
of the grand challenges in condensed matter theory and has resulted in
numerous controversies in the context of the cuprates.

Iron pnictides show very high resistivity and very large uniform
susceptibility in the normal state \cite{Kamihara:2008}.
Superconductivity emerges from a state of matter with highly enhanced
Pauli like type of susceptibility \cite{XRayNew}.
As a function of F$^-$ doping, the susceptibility is peaked around 5\%
doping reaching a value 25 times bigger than Pauli susceptibility
given by LDA \cite{XRayNew,Singh}.
In the parent compound, the resistivity exhibits a peak at 150$\,$K
\cite{Kamihara:2008} followed by a sharp drop, which is due to a
structural transition from tetragonal (space group P4/nmm) to
orthorhombic structure (space group Cmma) \cite{neutrons,XRayNew}
followed by a spin density wave transition \cite{SDW,XRayOld} at lower
temperature.
Only a 5\% doping completely suppress the specific heat anomaly
\cite{Cv}, while resistivity still shows a very steep drop below a
characteristic temperature $T^*$ \cite{Cv,QCP}. While there are
suggestions that the drop of resistivity is due to opening of a
pseudogap and proximity to a quantum critical point in samarium
compound \cite{QCP}, other measurements in lanthanum compound seem to
suggest less exotic and more Fermi liquid-like resistivity at 10\%
doping level\cite{Okridge,Hu}. Here we will show that the steep drop
of resistivity in doped compounds can be understood on the basis of
incoherence-coherence crossover, which is naturally accompanied by
gradual screening of the magnetic moments, leading to enhanced
magnetic susceptibility, and crossover from Currie-Weiss to Pauli type
susceptibility.

Theoretically, electron-phonon coupling is not sufficient to explain
superconductivity in LaO$_{1-x}$F$_{x}$FeAs \cite{Haule,Golubov} and
spin and orbital fluctuations need to be considered.  The weak
coupling approach based on RPA approximation has been carried out by
several groups for a simplified two band model containing two
orbitals, $d_{xz}$ and $d_{yz}$ \cite{RPA1,BestRPA,Scalapino3}.
The nature of the superconducting state and the pairing symmetry is
apparently very sensitively to the parameters of the model, and
therefore it is very desirable to reliably determine parameters of a
microscopic Hamiltonian, which is the purpose of this article.

The electronic structure of the LaO$_{1-x}$F$_{x}$FeAs compound was
studied by the density functional theory (DFT) \cite{Singh} and the
dynamical mean field theory (DMFT) \cite{Haule}.  DFT calculations
show that this material has a well separated set of bands with mostly
$d$ character in the interval between $-2\,$eV and $2\,$eV around the
Fermi level \cite{Haule}.  The parent compound has five Fermi surface
sheets \cite{Singh}, two electron cylinders, centered around the zone
edge (M-A line), and three hole pockets around the zone center
($\Gamma$-Z line).  Upon doping with electrons, the hole-like pocket
quickly disappear. The nesting wave vector between the hole and
electron pocket promotes the spin density wave instability which was
actually observed in experiment \cite{XRayOld}. Notice however that
the Bragg peak in neutron experiments remains commensurate and does
not change with doping, pointing to the inadequacy of the weak
coupling spin density wave scenario.

\begin{figure}[!ht]
\centering{
  \includegraphics[width=0.7\linewidth]{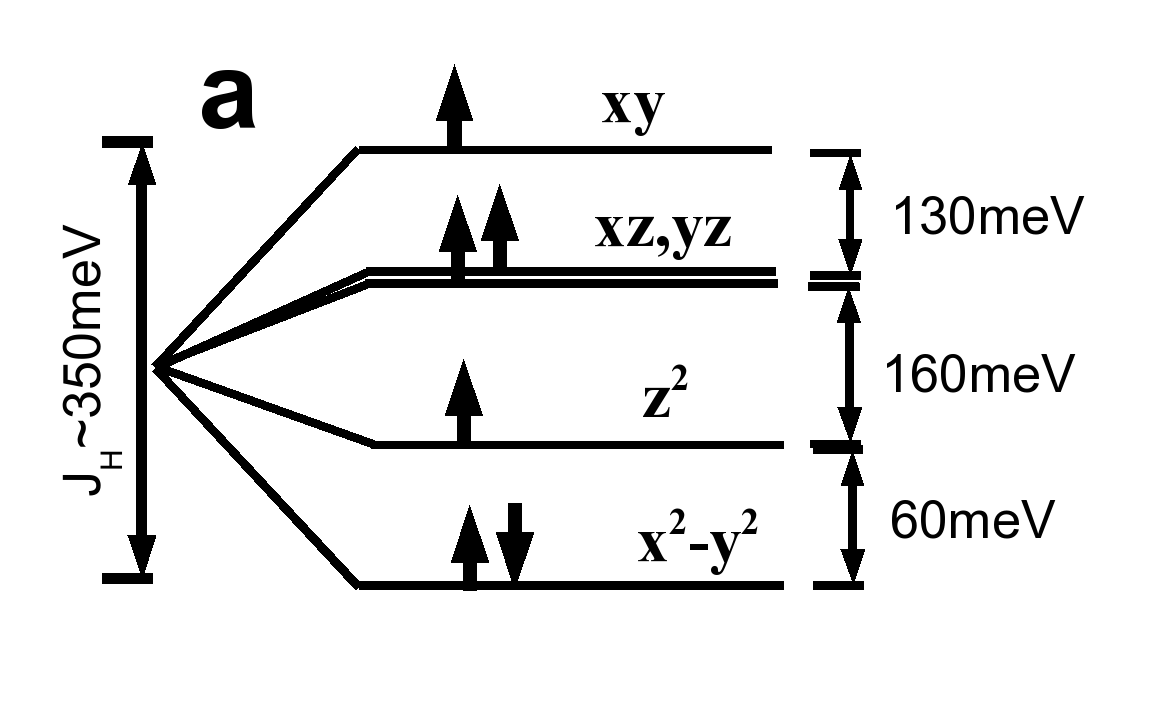}
  \includegraphics[width=0.7\linewidth]{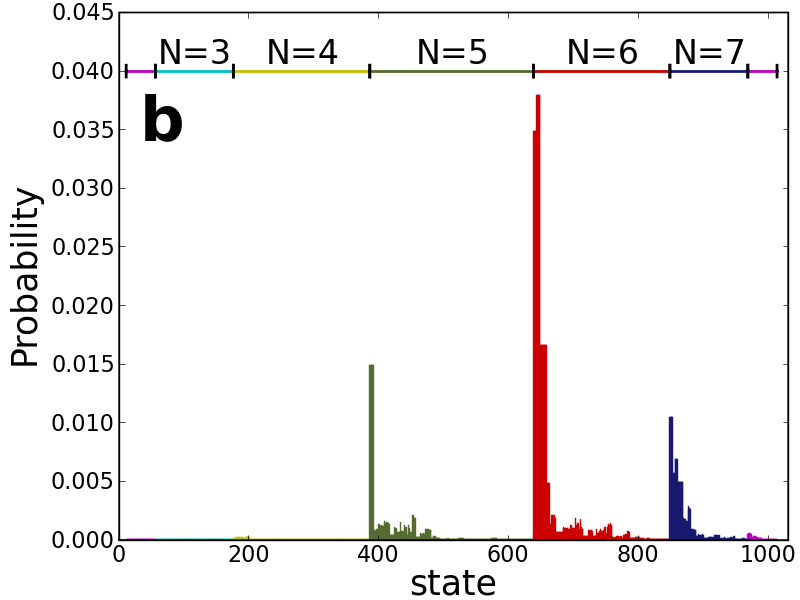}
  }
  \caption{a) Sketch of the orbital levels in the downfolded LDA
    Hamiltonian for LaOFeAs, keeping five d-orbitals (see appendix
    \ref{AppA}).  The levels are obtained by $h_{\alpha}=\sum_\vk
    H^{LDA}_{\vk,\alpha\alpha}$, where $\alpha$ is the orbital index.
    The Hund's coupling in this compound becomes relevant when its
    strength is comparable to the total splitting $J\sim 350\,$meV.
    The coordinate system is chosen such that $x$ and $y$ axis point
    from Fe atom towards its the nearest neighbor Fe atoms.  b)
    Probability for each iron $3d$ atomic state in DMFT calculation
    for LaOFeAs at $T=116\,$K and $J_{Hund}=0.4\,$eV. There are 1024
    atomic states in Fe-3d shell. Within sector with constant
    occupancy, states are sorted by increasing atomic energy).  }
\label{Probab}
\end{figure}
Fig.~\ref{Probab} sketches the crystal field levels, as obtained by
the Local Density Approximation (LDA), when the localized Wannier
orbitals are constructed for Fe-$3d$ states. Notice that the
tetragonal splittings are comparable to the t2g-eg splittings, leaving
only two degenerate orbitals, i.e., $xz$ and $yz$. Notice however,
that the bandwidth of these orbitals, as shown in Fig.3 of
Ref.~\cite{Haule}, is of the order of 3$\,$eV, which is one order of
magnitude larger than the crystal field splittings.

A naive atomic picture suggests a large magnetic moment on the Fe atom
($S=2$), and consequently almost classical moments, which can not
condense by Copper pairing, but would rather order magnetically, just
like they do in pure iron. 
Due to degeneracy of $xz$ and $yz$ orbital, even an infinitesimal
Hund's rule coupling would therefore lead to a spin state of at least
$S=1$.

Dynamical mean field studies \cite{Haule}, showed that the
correlations in the parent compound are strong enough to make this
material a bad metal with a large electronic scattering rate at, and
below room temperature, as seen in the momentum resolved photoemission
spectra, and the absence of a Drude peak in the optical conductivity.
However, it was argued in Ref.~\cite{Haule}, that the
correlations are {\it not } strong enough to open the Mott gap in the
parent compound, in contrast to the hole doped cuprates, which are
known to be doped Mott insulators. The DMFT study determined LaOFeAs,
to be a strongly correlated metal, which is not well described by
either atomic physics, nor band theory. Furthermore, it was shown that
all the five iron $d$ orbitals have appreciable one particle density
of states within $0.5\,$eV of the Fermi level, and therefore a
realistic minimal model for these materials should contain all five
$d$ iron orbitals.

In this article, we use the LDA+DMFT method \cite{review} in
combination with the continuous time Quantum Monte Carlo method
\cite{CTQMC}, to investigate the transport and thermodynamic
properties of the 10\% doped LaO$_{1-x}$F$_x$FeAs compound. More
details of the method are given in Ref.~\cite{Haule,review}.

Our goal is to understand why a material with high orbital degeneracy,
which naively would have expected to be an excellent metal
\cite{note1}, exhibits strong correlation effects, and what determines
the crossover from the incoherent regime to the coherent Fermi liquid
state.  We show that the Hund's rule coupling $J_{Hund}$ dramatically
reduces the coherence scale $T^*$ in these compounds, and promotes the
highly incoherent metallic state.

In physical terms, correlated quasiparticles are spin one half objects
that have to be built from atomic states, which are dominantly $S=2$
and $S=1$, in the presence of Hund's rule coupling.  As a result, the
overlap of the quasiparticle and the bare electronic states is very
small.

To understand the nature of the metallic incoherent state, it is
useful to study the probability that an electron in this compound is
found in any of the atomic states of iron $3d$ orbital. This is plotted
in the atomic histogram in Fig.~\ref{Probab} introduced in
Ref.~\cite{ctqmc}.  Even the most probable atomic states have
probability of only a few percent, hence a naive atomic limit is
qualitatively wrong for this compound.
The atomic histogram, characterizes the nature of the ground state of
the material. To investigate the coherence incoherence crossover, we
need to probe the excitation spectra, through the transport and the
thermodynamic probes.

\begin{figure}[!ht]
\centering{
  \includegraphics[width=0.7\linewidth]{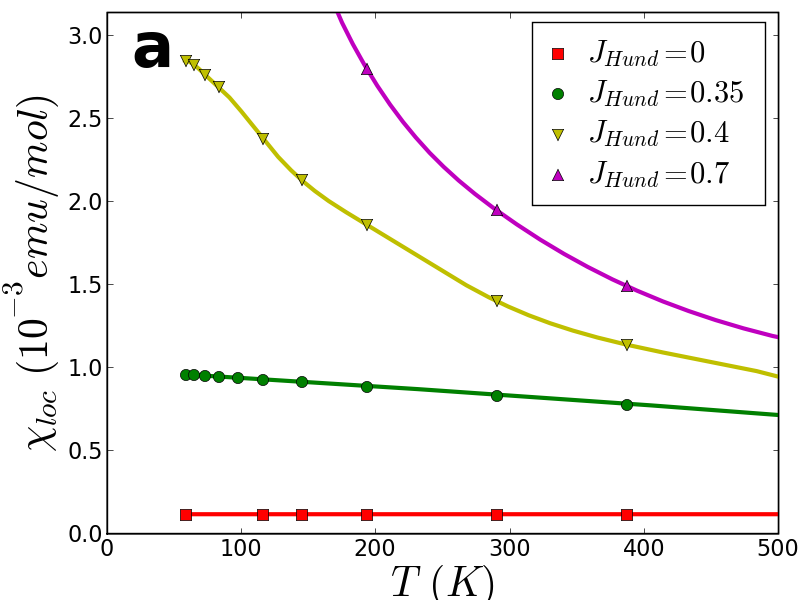}
  \includegraphics[width=0.7\linewidth]{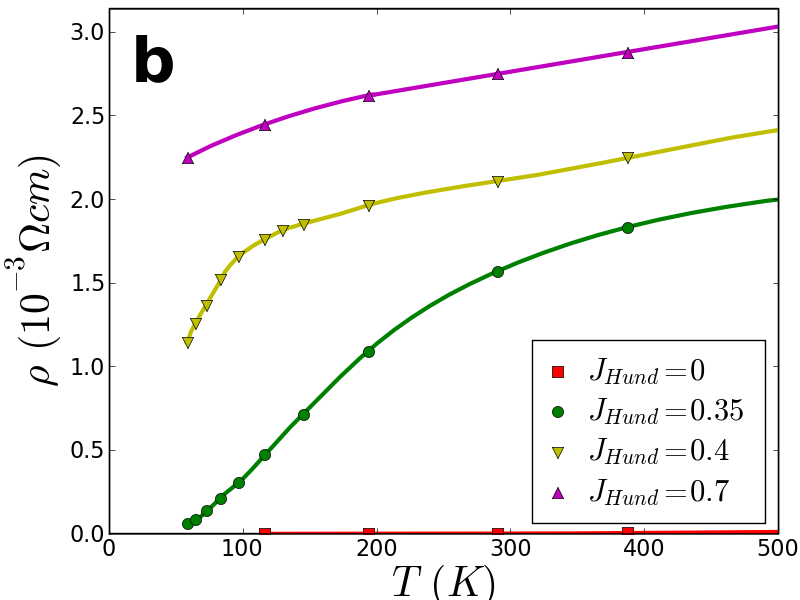}
  }
  \caption{ a) The local susceptibility versus temperature for few
    different Hund's coupling strengths.  b) Resistivity versus
    temperature for the same Hund's coupling strengths. }
\label{ChiRho}
\end{figure}
We first turn to the magnetic susceptibility which is the most direct
probe of the spin response:
\begin{equation}
\chi = \frac{(g\mu_B)^2 N_A}{k_B}\int_0^\beta d\tau \langle S_z(\tau)S_z(0)\rangle.
\end{equation}
Here $S_z$ is the spin of the iron atom, $g\mu_B/2$ is a magnetic
moment of a free electron, $N_A$ is the Avogadro number, and $\beta$
is the inverse temperature.  In the absence of screening (static
$S_z$), susceptibility should follow the Curie-Weiss law
$\chi=\frac{(g\mu_B)^2 N_A S(S+1)}{3 k_B T}$. In the opposite itinerant
limit, the spin susceptibility exhibits Pauli response $\chi=\mu_B^2
N_A D_0$, where $D_0$ is the density of state at zero frequency. In
Fig.~\ref{ChiRho}a, we show the local spin susceptibility for a few
values of Hund's coupling $J_{Hund}$ in 10\% doped compound.

In the absence of $J_{Hund}$, the susceptibility obeys the Pauli law, and
the mass enhancement due to correlations is negligible.
On the other hand, at large $J_{Hund}=0.7\,$eV, the system remains in the
local moment regime down to lowest temperature we reached (50$\,$K).
For some intermediate values of Hund's coupling $J_{Hund}\sim 0.35-0.4\,$eV,
the system evolves from an incoherent metal at high temperature, to
the coherent Fermi liquid at low temperature.

The metallicity of the compound always leads to screening of the
electron magnetic moments, which eventually disappear at sufficiently
low temperature and frequencies, but are present at high temperature
and high energies.  The scale at which the screening of the local
moment takes place, however, is a very sensitive function of of the
strength of the Hund's coupling, being very large for $J_{Hund}=0$ and
unobservably small for $J_{Hund}=0.7$ in the temperature range considered.
This is one of the main messages of this article.

To shed light into the coherence-incoherence crossover, we computed
the resistivity due to electron-electron interactions, given by
\begin{eqnarray}
  \label{eq:A_coefficients2}
  \frac{1}{\rho} = \frac{\pi e^2}{V_0 \hbar}
  \int{d\omega}\left(-\frac{df}{d\omega}\right)
  \sum_\vk\trace[v_\vk(\omega)\rho_\vk(\omega)v_\vk(\omega)\rho_\vk(\omega)]
\end{eqnarray}
where $v_\vk$ are velocities of electrons, $\rho_\vk$ is the spectral
density of the electrons, and $V_0$ is the volume of the unit cell.
The resistivity in Fig.~\ref{ChiRho}b demonstrates the crossover even
clearly. The resistivity is approximately linear with modest slope
above the coherence crossover temperature, and it exhibits a steep
drop around the coherence temperature. It is important to stress that
the drop of resistivity here is not due to proximity to a quantum
critical point \cite{QCP}, or due to proximity to spin density wave
\cite{SDW}, but because the electrons, being rather localized at high
temperature, start to form a coherent quasiparticle bands with the
underlying Fermi surface.

In most compounds, the strength of the correlations is controlled by
the ratio of the diagonal on-site Coulomb interaction $U$ to the
bandwidth, while Hund's coupling usually plays a sub-leading role.  We
find that in this compound, a crucial role is play by the Hund's
coupling.  This point was first noticed in the context of a two band
Hubbard model calculation at half-filling \cite{Pruschke}, and it was
shown that the coherence scale is exponentially suppressed when Hund's
coupling is taken into account.  Moreover, it was emphasized that the
non-rotationally invariant Hund's coupling, usually assumed in
combination with traditional quantum Monte Carlo methods
\cite{Anisimov}, can lead to a qualitatively and quantitatively wrong
results. Indeed, we have found that the absence of the rotational
invariant Hund's coupling leads to a first order phase transition as a
function of $J_{Hund}$, between a Fermi liquid and non-Fermi liquid
state at zero temperature.
On the other hand, the rotational invariance of the Coulomb
interaction recovers Fermi liquid at zero temperature, although the
coherence scale can become very small, as seen in Fig.~\ref{ChiRho}b
for $J_{Hund}=0.7$.

It is interesting to note that the atomic $J_{Hund}$ in iron is
approximately $\sim 1.2\,$eV. In many materials with correlated $f$
orbitals, the screening of $J_{Hund}$ in solid is well accounted for
by 20\% reduction from its atomic value. In LaOFeAs compound, only
30\% of the atomic Hund's coupling gives the best agreement with
experiment for resistivity and susceptibility.

\begin{figure}[!ht]
\centering{
  \includegraphics[width=0.7\linewidth]{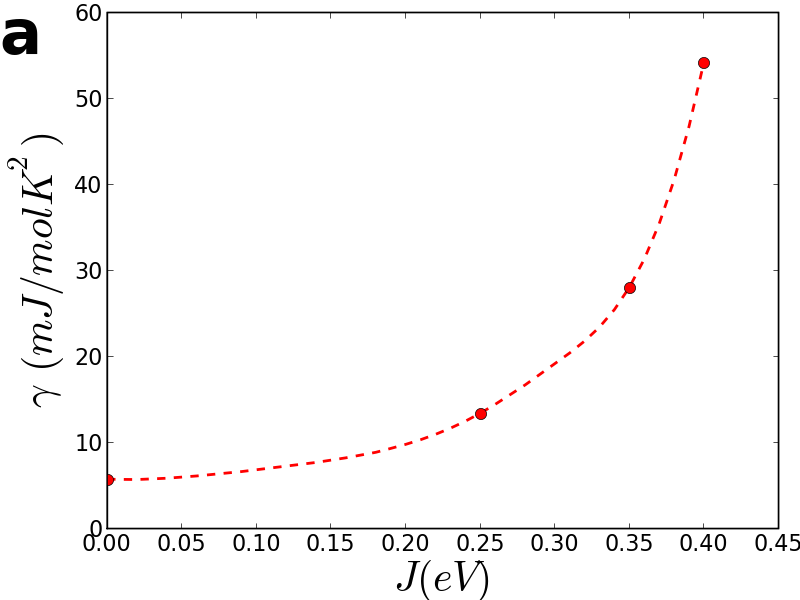}
  \includegraphics[width=0.7\linewidth]{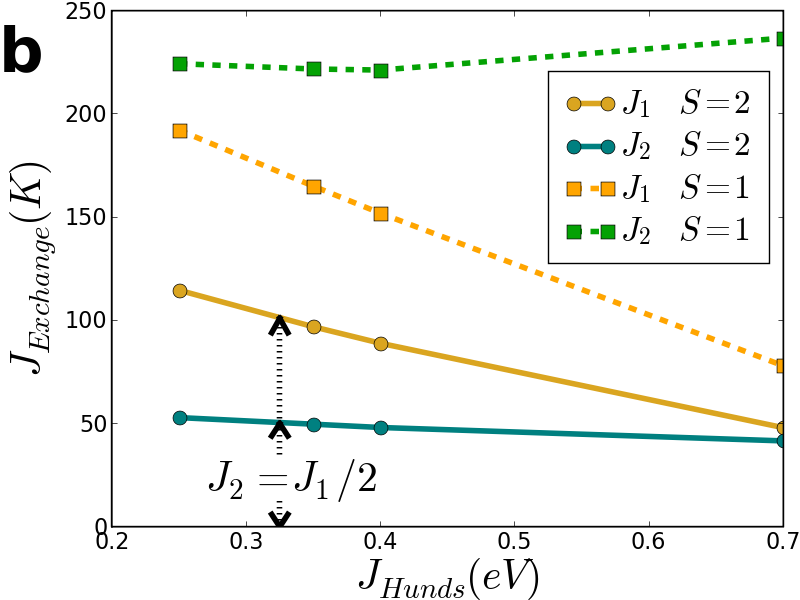}
  }
  \caption{ a) Low temperature Fermi liquid specific heat coefficient
    $\gamma$ versus Hund's coupling strength. The effective mass is
    exponentially enhanced by the Hund's coupling. The coherence scale
    is consequentially exponentially suppressed by the Hund's coupling.
    b) Exchange constants, as obtained from the second order
    perturbation theory around the atomic limit, considering either
    S=2,N=6 or S=1,N=6 for the ground state. Positive values for the
    exchange constants correspond to antiferromagnetic couplings.  The
    rest of the exchange constants are small, for example, $J_3\sim
    1.8\,$K and $J_z\sim 0.2\,$K.  }
\label{CvEx}
\end{figure}
To elucidate the sensitivity of mass enhancement with varying Hund's
coupling, we show low temperature specific heat in Fig.~\ref{CvEx}.
For small $J_{Hund}$, we find specific heat value very close to the
LDA prediction $\gamma\sim 6 mJ/(mol K^2)$. For $J_{Hund}=0.35\,$eV, a
value which gives resistivity in very favorable agreement with
experiments, requires the enhancement of the specific heat for factor
of 5 to $30\,mJ/(mol K^2)$. This is large enhancement and should be
easy to measurable in future experiments, and is a key prediction of
the theory.  This would be a strong test that a local theory, as the
one described in this paper, correctly predicts the physical
properties of this material, since would allow to differentiate it
from paramagnon or itinerant spin density wave theories, where the
enhancement of the static susceptibility is much larger than the
enhancement of the specific heat.
At this point, reliable data for the normal state specific heat of the
doped compound is not available down to sufficiently low temperature
to reliably extract the coefficient $\gamma$. It would be interesting
to measure it in the slightly doped compound ($5\%$), where the
superconducting temperature is low enough, and susceptibility shows
the strongest enhancement.

The full computation of the momentum dependent spin response within
LDA+DMFT is still a formidable task. However one can gain insights
into the momentum dependence of $\chi$ by making a simplifying
assumptions, that near the Mott transition $\chi(\vq)^{-1} = a +
J_\vq$, where $J_\vq$ is the Fourier transform of the exchange
constants, and $a$ is a local but frequency dependent object.
 
We computed the first neighbor $J_1$ and next-nearest neighbor $J_2$
exchange constants as a function of Hund's coupling.  For this
purpose, we first constructed Wannier orbitals from iron $3d$ bands
(see appendix \ref{AppA}), we exactly diagonalized the iron atom, and
perform the second order perturbation theory, with respect to
iron-iron hopping. This perturbation is carried out around the atomic
ground state of the $S=2$ and $S=1$ sector.

Fig ~\ref{CvEx}b shows the exchange constants as a function of the
Hund's rule coupling.  This provides complementary information to the
exchange constants computed around the band limit, which were recently
determined by LDA \cite{Yidrim,Js1,Sergej}, and are typically larger
in magnitude than those obtained here. 
The exchange constants are supportive of the actual ordering in the
parent compound, determined recently by the neutron scattering
experiment\cite{neutrons}, if $J_2>J_1/2$. For reasonable Hund's
coupling, both $S=1$ and $S=2$ ground states respect this inequality.
If we select the $S=1$ state for the ground state, the next-nearest
$J_2$ is much bigger than $J_1$.
For the actual ground state of the atom ($S=2$),
%
%
and realistic $J_{Hund}\sim 0.35$, the
value of $J_2$ is close to $J_2 >= J_1/2$. It is interesting to note
that according to a well known result \cite{Chandra}, this particular
choice of exchange constants results in large magnetic fluctuations
due to frustration. At temperature above the spin density wave
transition, there can also be a second order Ising phase transition,
at which the two sublattice magnetizations of the two interpenetrating
Neel sublattices, lock together. According to Ref.~\cite{Chandra}, the
Ising transition would occur around $T_c \sim J_2 S^2 \sim 200K$.

The large value of the next nearest neighbor exchange constant
demonstrates that even in the atomic limit, the system is highly
frustrated, as pointed out in Ref.~\cite{Yidrim,Elihu}, which gives us
more confidence that the single site DMFT description, employed in
this article, is adequate, and the momentum dependent self-energy, as
obtained by the cluster extensions of DMFT \cite{review}, will not
qualitatively change the physics.

From the above provided evidence, LaO$_{1-x}$F$_{x}$FeAs system fits
nicely within the general rubric of strongly correlated
superconductivity.  The hallmark of this phenomena, is the appearance
of a incoherent normal state, above $T_c$.
In the classic Bardeen-Cooper-Schrieffer theory of superconductivity,
the effective Fermi energy is much larger than the pairing
interaction, resulting in small critical temperatures.  Strong
correlations play the role of reducing the the kinetic energy,
allowing the pairing interactions to lead to appreciable critical
temperatures \cite{Capone}.

While these are common themes shared by all the high $T_c$ materials,
the mechanisms by which different systems manage to renormalize down
their kinetic energies, and the details of the competing states that
appear as a result, are different in different classes of materials.

Hole doped cuprate superconductors can be modeled as a single-band
materials, with Coulomb correlation slightly above the Mott transition
(in DMFT terminology the effective $U$, which arises from the charge
transfer energy, is above $U_{c2}$), and indeed maximal $T_c$ is
achieved when $U$ is in the vicinity of critical $U_{c2}$.  

We have argued in our earlier publication \cite{Haule}, that the
LaOFeAs is an intrinsically multiorbital system, which is slightly {\it
  below} $U_{c2}$.  The mechanism which is responsible for the
incoherence of the normal state and the renormalization of the kinetic
energy is the Hund's rule $J_{Hund}$, which we estimate here to be
$J_{Hund}=0.35-0.4$ eV.  Competing states in this material are magnetic
states, that have been predicted in Ref.~\cite{SDW} and since observed
by neutron scattering experiments \cite{neutrons,XRayOld,XRayNew}.

The incoherent normal states, is then eliminated by either a magnetic
or a superconducting state, and the study of the competition between
these mechanism in the multiorbital framework, deserves further
attention.

\appendix

\section{The tight binding Hamiltonian for the Fe-As planes}
\label{AppA}

\begin{figure}[!ht]
\centering{
  \includegraphics[width=0.85\linewidth]{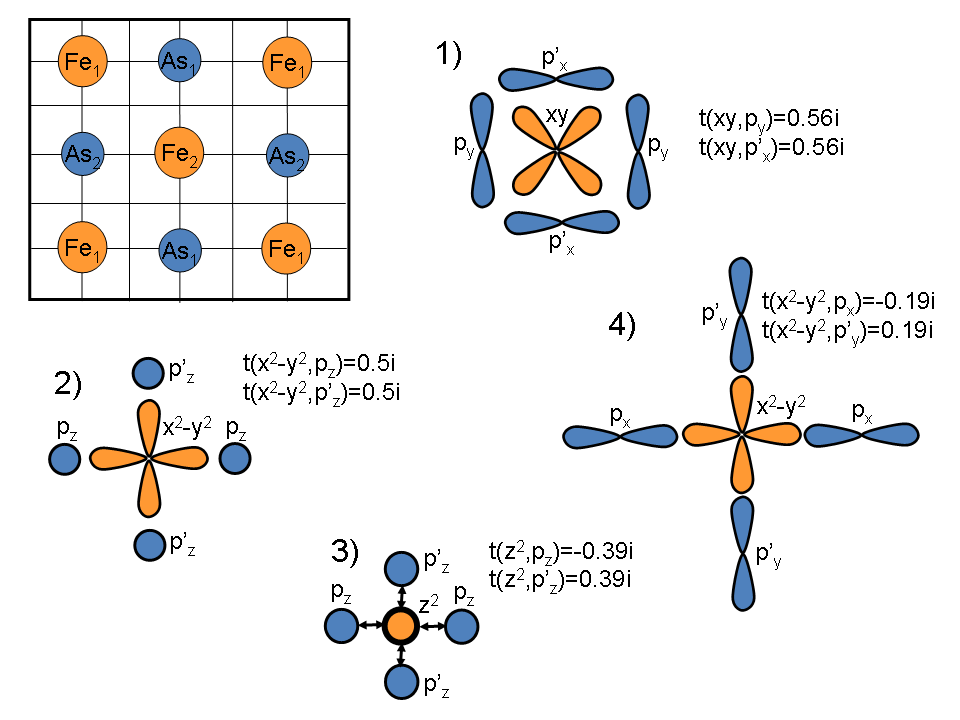}
  }
  \caption{left top) choice of the unit cell in the FeAs plane. As$_1$
  lays above the Fe plane, and As$_2$ below it. 1-4) Sketch of some
  largest hoppings and their values.
}
\label{F1a}
\end{figure}

The Fe-As planes in LaFeAsO can be described by a tight-binding
Hamiltonian containing Fe and As atoms, with two Fe and two As per
unit cell (see Fig.~\ref{F1a} for drawing of the unit cell containing
Fe$_1$, Fe$_2$, As$_1$ and As$_2$). We obtained the tight-binding
parameterization by downfolding the Local Density Approximation
first-principles results, which are expressed in a well localized
Muffin Tin Orbital (MTO) base. The localized nature of the MTO base
makes the projection of Kohn-Sham orbitals to a localized orbital
basis set simple, and downfolding procedure more constrained and well
defined.

\begin{figure}[!ht]
\centering{
  \includegraphics[width=0.85\linewidth]{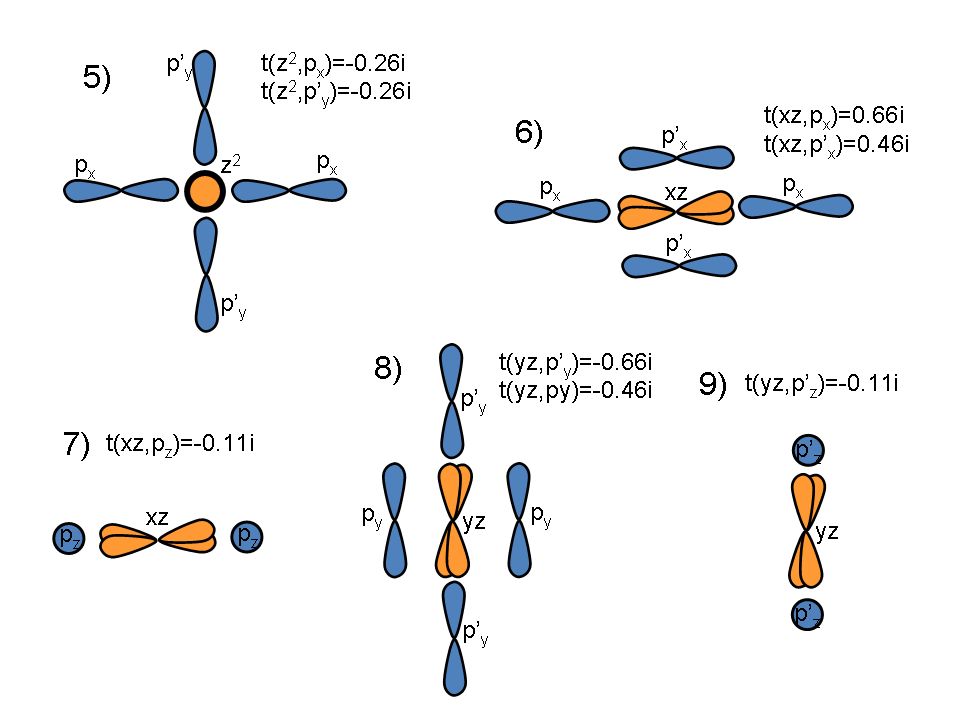}
  }
  \caption{Sketch of a few more relevant Fe-As hoppings.}
\label{F1b}
\end{figure}

The few largest hoppings are sketched in Fig.~\ref{F1a} and
\ref{F1b}. All energies and hoppings are in units of eV.
The on-site energies of all orbitals are tabulated in table~\ref{T1}
and most of the hoppings are tabulated in tables~\ref{T2} and
\ref{T3}. Note that orbitals are expressed in the coordinate system of
the four atoms unit cell, shown in Fig.~\ref{F1a}. The complete set of
tight binding hoppings is available to download \cite{calc}. For
comparison we also include the hoppings for a closely related compound
LaFePO, which shows considerably less correlation effects due to
somewhat larger crystal field splittings, which renormalize down the
effective Hund's coupling.

\begin{table}
\begin{center}
\begin{tabular}{lccc}
$\textbf{E}-onsite$    & & \textbf{LaFeAsO} & \textbf{LaFePO} \\
\hline
$E[x^2-y^2]$     &       &   -0.510 & -0.654 \\  
$E[yz]$     &       &   -0.535 & -0.656 \\  
$E[z^2]$     &       &   -0.564 & -0.635 \\  
$E[xy]$     &       &   -0.650 & -0.692 \\  
$E[p_z]$     &       &   -2.144 & -1.990 \\  
$E[p_x]$     &       &   -2.440 & -2.358 \\  
\end{tabular}
\end{center}
\caption{On-site energies for both LaFeAsO and LaFePO in the model
  with both Fe-d and As-p orbitals.}
\label{T1}
\end{table}
\begin{table}[h]
\begin{center}
\begin{tabular}{lcrr}
\textbf{hopping}    & \textbf{figure} & \textbf{LaFeAsO} & \textbf{LaFePO} \\
\hline
$xy-p_y$   &  (1)  &   0.56*i &  0.64*i\\
$xy-p_x'$  &  (1)  &   0.56*i &  0.64*i\\  
\hline
$x^2-pz$   &  (2)  &   0.5*i  &  0.58*i\\  
$x^2-p_z'$  &  (2)  &   0.5*i  &  0.58*i\\  
$x^2,p_x$   &  (4)  &  -0.19*i & -0.33*i\\  
$x^2,p_y'$  &  (4)  &   0.19*i &  0.33*i\\  
\hline
$z^2,p_z$   &  (3)  &  -0.39*i & -0.45*i\\  
$z^2,p_z'$  &  (3)  &   0.39*i & -0.45*i\\  
$z^2,p_x$   &  (5)  &  -0.26*i & -0.22*i\\  
$z^2,p_y'$  &  (5)  &  -0.26*i & -0.22*i\\  
\hline
$xz,p_x$   &  (6)  &   0.66*i &  0.76*i\\  
$xz,p_x'$  &  (6)  &   0.46*i &  0.51*i\\  
$xz,p_z$   &  (7)  &  -0.11*i &  0.0   \\  
\hline
$yz,p_y$   &  (8)  &  -0.46*i & -0.51*i\\  
$yz,p_y'$  &  (8)  &  -0.66*i & -0.77*i\\  
$yz,p_z'$  &  (9)  &  -0.11*i &  0.0   \\  
\hline
\end{tabular}
\end{center}
\caption{
  The few largest hopping matrix elements in the model of Fe-d and As-p
  orbitals. Each Fe atom is surrounded by two inequivalent As-atoms. 
The two different As orbitals are sketched in
  Fig.~\ref{F1a},\ref{F1b} and are denoted by $p$ and $p'$.
}
\label{T2}
\end{table}
\begin{table}[h]
\begin{center}
\begin{tabular}{lrr}
\textbf{hopping}   & \textbf{LaFeAsO} & \textbf{LaFePO} \\
\hline
$p_z,p_z'$  &    0.29   &  0.27  \\  
$p_x,p_x'$  &    0.16   &  0.21  \\  
$p_y,p_y'$  &    0.16   &  0.21  \\  
$p_x,p_z'$  &   -0.25   & -0.27  \\  
$p_y,p_z'$  &    0.25   &  0.27  \\  
$p_z,p_x'$  &   -0.25   & -0.27  \\  
$p_z,p_y'$  &    0.25   &  0.27  \\  
$p_x,p_y'$  &   -0.15   & -0.18  \\  
$p_y,p_x'$  &   -0.15   & -0.18  \\  
\hline
$xy,xy'$	&   -0.30  & -0.35  \\  
$xy,{z^2}'$     &    0.21  &  0.24  \\  
$z^2,xy'$	&    0.21  &  0.24  \\  
$x^2,{x^2}'$	&    0.19  &  0.18  \\  
\hline
$yz,xz'$	&    0.15  &  0.16  \\  
$xz,yz'$        &    0.15  &  0.16  \\  
\hline
$yz,{x^2}'$	&    0.09  &  0.12  \\  
$xz,{x^2}'$	&    0.09  &  0.12  \\  
$x^2,yz'$       &    0.09  &  0.12  \\  
$x^2,xz'$	&    0.09  &  0.12  \\  
\hline
$z^2,{z^2}'$    &   -0.07  & -0.06  \\  
$yz,yz'$        &    0.06  &  0.07  \\  
$yz,xy'$	&    0.06  &  0.07  \\  
$xz,xz'$	&    0.06  &  0.07  \\  
$xz,xy'$	&   -0.06  & -0.07  \\  
$xy,yz'$        &    0.06  &  0.07  \\  
$xy,xz'$	&   -0.06  & -0.07  \\  
\hline
$yz,{z^2}'$     &   -0.02  & -0.02  \\  
$xz,{z^2}'$     &    0.02  &  0.02  \\  
$z^2,yz'$       &   -0.02  & -0.02  \\  
$z^2,xz'$	&    0.02  &  0.02  \\  
\hline
\end{tabular}
\end{center}
\caption{Continued from table \ref{T2}, a few more large hopping matrix
  elements of the Fe-d,As-p model.}
\label{T3}
\end{table}

\newpage

\begin{figure}[!ht]
\centering{
  \includegraphics[width=0.99\linewidth]{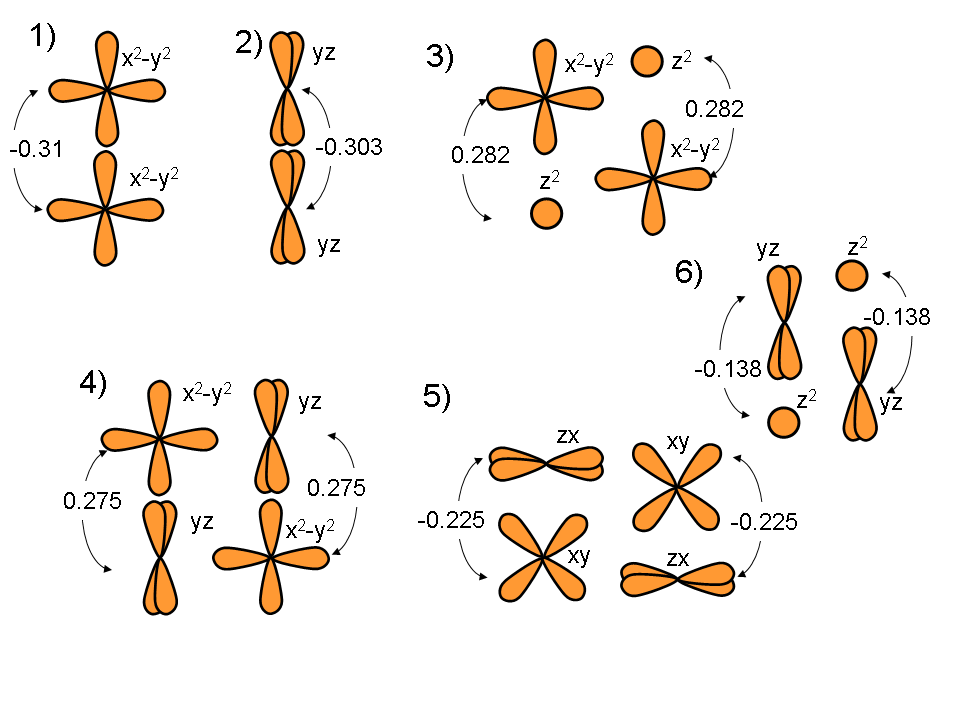}
  } 
  \caption{The sketch of the largest Fe-Fe hoppings in the effective five
    band Hubbard model of irons only.}
\label{F3}
\end{figure}

Further downfolding to the 5 band model of Fe is possible. The on-site
energies and the few largest hoppings of the corresponding
multiorbital Hubbard model are tabulated in table 3. The sketch of the
most relevant hoppings is shown in Fig.~\ref{F3}. Note that the
coordinate system here is chosen differently than above. Here we use
the 45$^\circ$ rotated coordinate system, which corresponds to usual square lattice of
irons.

\begin{table}[h]
\begin{center}
\begin{tabular}{lcrr}
\textbf{hops} & \textbf{fig}  & \textbf{LaFeAsO} & \textbf{LaFePO} \\
\hline
$E[xy]$     &     &  0.096 &  0.244 \\
$E[yz]$     &     & -0.034 & -0.054 \\
$E[z^2]$     &     & -0.189 & -0.201 \\
$E[x^2]$     &     & -0.251 & -0.162 \\
\hline                          
$x^2,{x^2}'$	& (1) & -0.310 & -0.343 \\
$yz,yz'$  & (2) & -0.303 & -0.356 \\
$x2,{z^2}'$  & (3) &  0.282 &  0.310 \\
$z2,{x^2}'$	& (3) &  0.282 &  0.310 \\
$x2,yz'$  & (4) &  0.275 &  0.356 \\
$yz,{x^2}'$	& (4) &  0.275 &  0.356 \\
$zx,xy'$	& (5) & -0.225 & -0.269 \\
$xy,zx'$	& (5) & -0.225 & -0.269 \\
$yz,{z^2}'$  & (6) & -0.138 & -0.104 \\
$z^2,yz'$  & (6) & -0.138 & -0.104 \\
$zx,zx'$	&     & -0.010 & -0.035 \\
$xy,xy'$	&     & -0.086 & -0.175 \\
$z^2,{z^2}'$  &     &  0.036 &  0.088 \\
$zx,{z^2}'$  &     &  0.004 &  0.003 \\
$z^2,zx'$	&     &  0.004 &  0.003 \\
$zx,{x^2}'$	&     & -0.003 &  0.006 \\
$x^2,zx'$	&     &  0.003 & -0.006 \\
\hline
\end{tabular}
\end{center}
\caption{The on-site energies and the most important hoppings in the
  five band model of Fe-d orbitals only.}
\label{T4}
\end{table}


\begin{thebibliography}{99}
\bibitem{note1} Technically, the critical value of the Hubbard $U$,
  which is needed to obtain a insulating state ($U_{c2}$) in an SU(N)
  symmetric model is known to scale linearly with orbital degeneracy
  as $N$.  For this reason, Ref.~\cite{Anisimov}v1, which ignored the
  Hund's rule coupling, concluded that these materials are non
  correlated metals.
\end{thebibliography}
\end{document}